\documentclass[twocolumn,showpacs,prb,superscriptaddress,twoside]{revtex4}

\usepackage{epsfig}

\newcommand{\bdm}{\begin{displaymath}}
\newcommand{\edm}{\end{displaymath}}
\newcommand{\benl}{\begin{equation}}
\newcommand{\be}[1]{\begin{equation}\label{#1}}
\newcommand{\ee}{\end{equation}}
\newcommand{\bea}{\begin{eqnarray}}
\newcommand{\eea}{\end{eqnarray}}
\newcommand{\ba}{\begin{array}}
\newcommand{\ea}{\end{array}}

\begin{document}

\preprint{PRE/BR/Rosini}

\title{Noise and diffusion of particles obeying asymmetric exclusion
  processes}

\author{Marcello Rosini} \email{marcello.rosini@unile.it}
  \affiliation{CNISM and Dipartimento di Ingegneria dell'Innovazione
  Universit\`a di Lecce via Arnesano s/n, 73100 Lecce, Italy}

\author{Lino Reggiani}
  \affiliation{CNISM and Dipartimento di Ingegneria dell'Innovazione
  Universit\`a di Lecce via Arnesano s/n, 73100 Lecce, Italy}
  \affiliation{CNR-INFM  National
  Nanotechnology Laboratory, via Arnesano s/n, 73100 Lecce, Italy}

\date{\today}

\begin{abstract}
  The  relation  between noise  and  Fick's  diffusion coefficient  in
  barrier limited    transport associated with  hopping   or tunneling
  mechanisms  of particles  obeying  the asymmetric  simple  exclusion
  processes (ASEP) is physically  assessed by Monte Carlo simulations.
  For a  closed ring  consisting of a  large  number  of barriers  the
  diffusion coefficient  is related  explicitly  to the  current noise
  thus revealing  the existence   of  a  generalized  Nyquist-Einstein
  relation.  Both diffusion and noise are confirmed to decrease as the
  square root   of the  number  of barriers  as   a consequence of the
  correlation induced by ASEP.  By  contrast, for an open linear chain
  of  barriers  the  diffusion coefficient  is found  to  be no longer
  related   to  current noise.   Here  diffusion  depends  on particle
  concentration but is independent of the number of barriers.
\end{abstract}


\maketitle


The inter-relations between noise and diffusion in charge transport is
a  pillar    of   non-equilibrium  kinetics since    the   theoretical
interpretation of the Brownian motion by  Einstein. \cite{ein05} For a
kinetic  described   within   a  continuous  transport    model, where
quasi-particles  undergo local  scattering  events between  stochastic
free flights,  the   inter-relation between  noise  and diffusion  was
investigated by a number of     theoretical approaches ranging    from
analytical  models  \cite{pri65,van71,gan79,kog96,kat04} to  numerical
solutions of  the  appropriate kinetic  equations.  \cite{reg88,shi01}
The case of a barrier  transport model, dominated by tunnelling and/or
hopping  processes,  is less    developed.   Here, noise   was  mostly
investigated  for the case   of single and multiple quantum  barriers.
\cite{dej95,dutisseuil96,bee97,ian97,bla00,ale01}  By contrast, a  few
seminal works  have tackled  the problem of  noise in  hopping systems
\cite{kor1,sverdlov01} and  that of diffusion   in both tunneling  and
hopping systems.    \cite{ignatov85,bry1,bry2,ros1} For the  case of a
very large number of barriers, the asymmetric simple exclusion process
(ASEP) has been widely used in the recent past  as a relevant physical
model       for    the    description           of     non-equilibrium
dynamics.\cite{der1,khorrami00,huang01,klumpp04,roshani05}   In   this
context,  two systems of basic  interest  are the closed  ring and the
open linear chain consisting of a set of  multiple barriers, which are
the prototypes   of  closed and  open   systems driven  by  hopping or
tunneling transport mechanisms.  Here,  diffusion was investigated  in 
by  analytical means, \cite{der1,der2} and current noise
with Monte Carlo simulations. \cite{kor1,sverdlov01}
However, the  attempt  to interrelate diffusion   and current noise in
these systems remains a largely  unexplored issue.  In particular, the
dependence or less of diffusion on the number  of barriers, as well as
the prediction  of a partial or  complete suppression of shot noise in
the  presence of a  large number  of  barriers are intriguing features
still     lacking     of     a        microscopic      interpretation.
\cite{kor1,sverdlov01,mend06}
The  aim of  the  present  work is to  address   this issue  by  first
principles    Monte Carlo   simulations.   Accordingly,  diffusion  is
obtained by the calculation of the  time evolution of the spreading in
space  of a particle ensemble and  current noise by the calculation of
the autocorrelation  function  of current fluctuations  as measured in
the outside  circuit.  The  main features  of diffusion  and noise and
their interrelation are   thus  quantitatively assessed on  a  kinetic
physical ground.

We  take a physical system  consisting  of a  number  $N_w$ of  hopping
sites, separated  by a constant  distance $l$,  whose total length  is
$L=N_wl$.  The tunneling rate between  two adjacent site is assumed to
have the    same value  $\Gamma$.   We then     consider alternatively
periodic-  (closed  ring) or open-boundary (linear  chain) conditions.
By   imposing  a current  determined  by  a transition  rate $\Gamma =
10^{13}$ s$^{-1}$ to a structure with $l= 3.2$  nm (which are taken as
plausible parameters  for a real case)  we evaluate current, diffusion,
and  noise making use  of  an ensemble Monte   Carlo simulator.  It is
convenient    to   define   the dimensionless  carrier   concentration
$\rho=\langle N\rangle / N_w$, where $\langle N\rangle$ is the average
number  of carrier inside the  sample.  We then introduce correlations
between  carriers   within  the ASEP   model  by  imposing  a  maximum
occupation number $\nu$ for each site.  In particular, when $\nu=1$ if
a site is  occupied by one carrier then  no other carrier can jump to
this site, thus  carriers are strongly correlated. When $\nu\to\infty$
a  site  can be occupied  by   an arbitrary number  of  carriers, thus
carriers are totally uncorrelated.
The instantaneous current $I(t)$ is calculated as \cite{var94}
\be{current}
I(t)=\frac{e}{L}\sum_{i=1}^{N(t)} v_i(t) = \frac{e}{L} N(t) v_d(t) \ ,
\ee
where $e$  is  the unit  charge, $N(t)$  the  instantaneous  number of
carriers inside the  structure, $v_i(t)$ the instantaneous velocity of
the $i$-th carrier, $v_d(t)$  the instantaneous drift velocity  of the
carrier  ensemble. Under steady state  $I(t)$ is a stochastic variable
that  accounts for fluctuations in  carrier  number and velocity.   In
particular, for our discrete system $v_i(t)=l\delta \xi_i / \delta t$,
with $\xi_i$ the position index of the $i$-th particle. \cite{ros1}
\par
For the (longitudinal) diffusion  coefficient $D$, following Fick's law we
make  use  of  its  definition   as   a  spatial  spreading  quantity,
\cite{reggiani85}
\be{cdiff}
D=\frac{1}{2}\frac{\delta\langle(\Delta z(t))^2\rangle}{\delta t}\ ,
\ee
where $\Delta z(t)$ is   the distance between the  final
and  the  initial  hopping sites,     brackets  mean average  over   a
statistical ensemble  (up to  $10^3$) of  identical systems,  and the
time derivative is   carried out in a   time domain shorter  than  the
transit time but sufficiently long to  extrapolate the long time limit
for which D is found to be independent of time.
\par
The spectral density   of current fluctuations at zero  frequency is
\cite{var94}
\be{SIdiv}
S_I = 4\int_0^{\infty} dt  \langle\delta I(0)\delta I(t)\rangle
    = S_I^{v_d} + S_I^{N} + S_I^{Nv_d} + S_I^{v_dN}\ ,
\ee
where $S_I^{v_d}$, $S_I^{N}$ and $(S_I^{Nv_d} + S_I^{v_dN})$
refer to   the  three contributions (drift velocity,  number and  
cross-correlations between them)  in which the total spectral
density   can   be decomposed.   With Monte  Carlo
simulations these terms can be calculated separately.
The Fano factor is
$\gamma = S_I/(2e \langle I \rangle)$.

%
%
\par

\begin{figure}[t]
\centering
\includegraphics*[height=4.5cm]{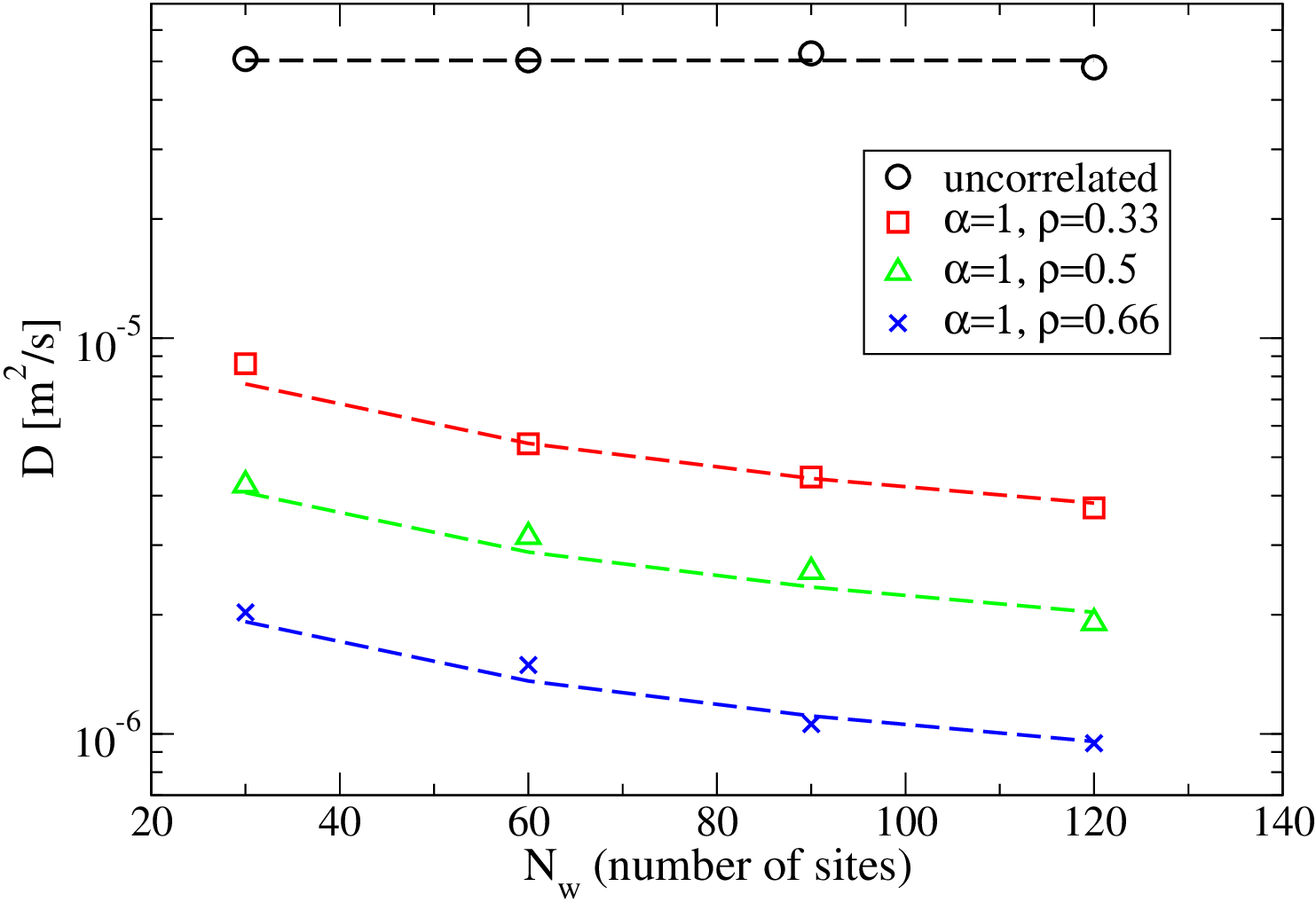}
\caption{  \label{f1}  (Color online) Cloded ring.  Comparison between
  the  analytical diffusion   coefficient in  Eqs.  (\ref{dASEPr}) and
  (\ref{dFREEr}) (dashed  lines) and   that obtained from  simulations
  (symbols)}
\end{figure}
\begin{figure}[t]
\centering
\includegraphics*[height=4.5cm]{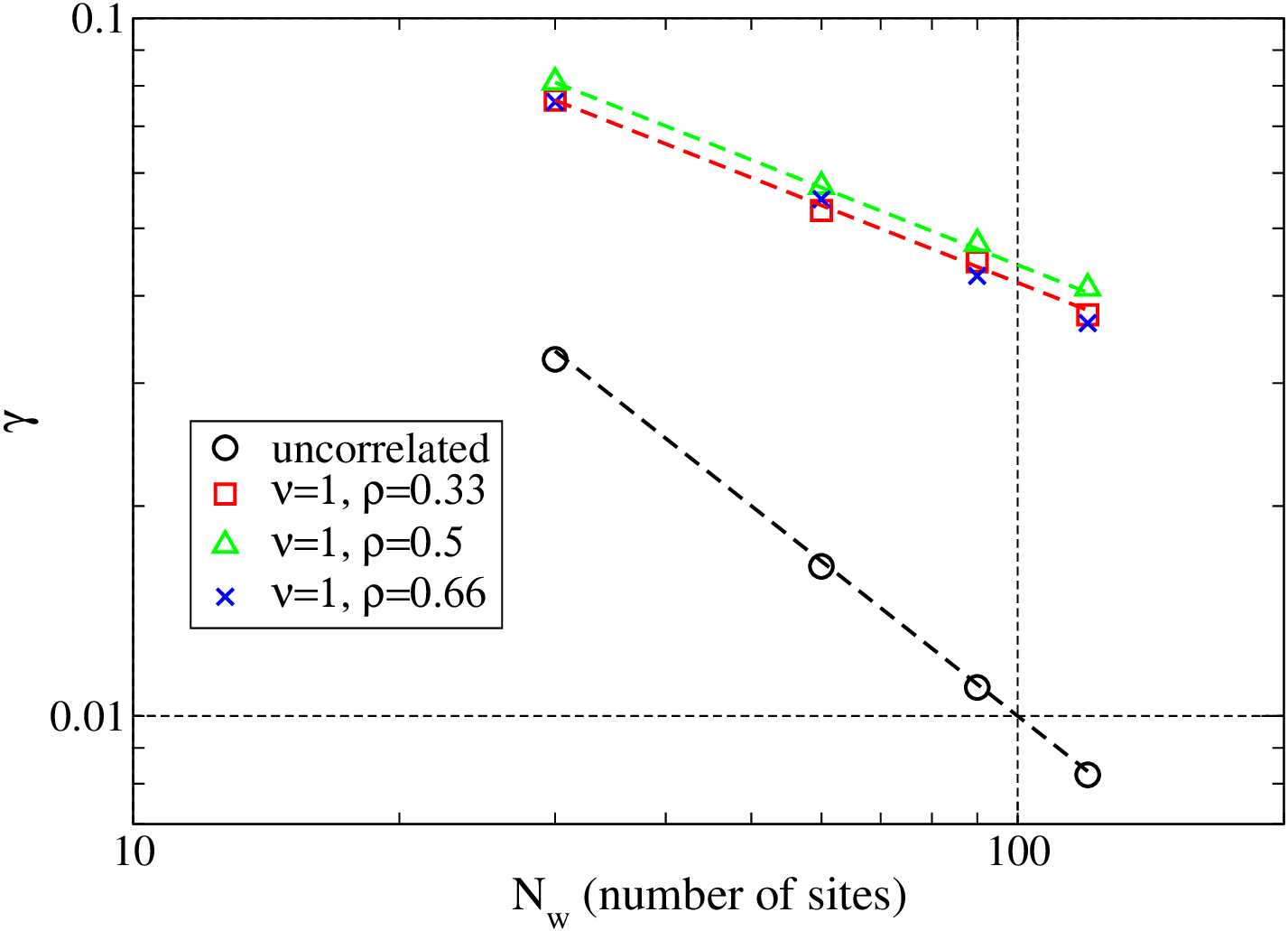}
\caption{ \label{f2}   (Color online) Closed  ring. Comparison between
  the analytical  Fano  factor in Eqs.  (\ref{gASEPr}), (\ref{gFREEr})
  (dashed lines) and that obtained from simulations (symbols).}
\end{figure}
{\it A. Closed ring.}  The  periodic  boundary conditions adequate  to
this structure  consists in  imposing that, for    a finite number  of
sites, the last site is directly connected to the first one by the 
same hopping rate $\Gamma$.
Let us consider the  case of correlated  carriers 
($\nu=1$).  For a
given   carrier concentration and  for  large  $N_w$ analytical theory
\cite{der1,kor1} gives the    following predictions: for  the  average
current
\be{iASEP}
\langle I\rangle_1 = e   \Gamma \rho (1- \rho)\ ,
\ee
and  for the diffusion coefficient  $\Delta_1$
(following [\onlinecite{der1}], with the subscript $1$ 
labeling the case of the  ring  geometry),
\be{dASEPr}
\Delta_1 = \frac{l^2  \Gamma}{2}  \frac{\sqrt{\pi}}{2}
  \frac{(1-\rho)^{\frac 3   2} }{ (\rho N_w)^{\frac  1  2}}\ .
\ee
[We suppose that in general  the value  of $\Delta_1$ differs from
that   of $D$ in Eq.(\ref{cdiff}).]   
\par 
The results  of the simulations  for the diffusion
coefficient   are  reported in   Fig.    \ref{f1}.   
Here,  the identity $\Delta_1  =  D$ is confirmed
for  all  the concentrations considered.
\par
For the  current noise  (in  this  case  due only  to velocity
fluctuations since the number of   carriers is rigorously constant  in
time) the simulations confirm the relation
\be{eq7}
S_I \equiv  S_I^{v_d}=\frac{4e^2}{l^2} \rho^2  \Delta_1 \ ,
\ee  
with the corresponding Fano factor (Fig. \ref{f2}), given by
\be{gASEPr}
\gamma=\frac{\sqrt{\pi}}{2}
\frac{\rho^{1/2} (1-\rho)^{1/2}}{N_w^{1/2}}\ .
\ee 
Equation
  (\ref{eq7}), by revealing a strict relation between  noise
and   diffusion,  takes    the  form  of  a   generalized
Nyquist-Einstein  relation  \cite{pri65,reg88,balandin02}.   
The reason of diffusion
and  noise suppression as  $1/\sqrt{N_w}$,
 confirmed by the simulations,
is  attributed to the strong  correlation  among carriers.  To support
this   interpretation, we considered   also  the case of  uncorrelated
carriers  (i.e.  in  the absence of  ASEP) where,  for a given carrier
concentration, analytical  theory gives the following predictions: for
the average current \cite{kor1}
\be{iFREE}
\langle I\rangle_0 = e   \Gamma \rho\ ,
\ee
for the  diffusion coefficient \cite{bry1,bry2}   
\be{dFREEr}
D_0=\frac{l^2 \Gamma}{2}\ ,
\ee
(with the  subscript $0$ labeling the  case of uncorrelated particles)
and for the current noise the standard Nyquist Einstein relation 
\cite{var94}
\be{eq11}
S_I \equiv S_I^{v_d} = \frac{4e^2}{l^2}  \rho^2 \frac{D_0}{N}\ ,
\ee
with the corresponding Fano factor
\be{gFREEr}
\gamma =\frac{1}{N_w}\ .
\ee
The result  of  simulations  confirms that,   in the absence  of  ASEP,
diffusion becomes independent of $N_w$  (see the curve uncorrelated in
Fig.    \ref{f1}), and the Fano factor   decreases as $1/N_w$ (see the
curve uncorrelated in Fig.  \ref{f2}).
In  all  the cases considered   here (even  when   $\nu=1$ so that the
non-passing  constraint is  accomplished)  the time  evolution of  the
variance  in  space of the carrier  ensemble   is found to  be linear,
contrarily to  the suggestion of  a sub-diffusive (and thus sublinear)
behavior. \cite{der1}  On the other  hand,
the suppression of  diffusion in the presence of  ASEP is  found to be
related to a subpoissian variance  of the number of scattering events,
which is found to    exhibit the expected   $1/\sqrt{N_w}$
behavior.
\par

\begin{figure}[t]
\centering
\includegraphics*[height=4.5cm]{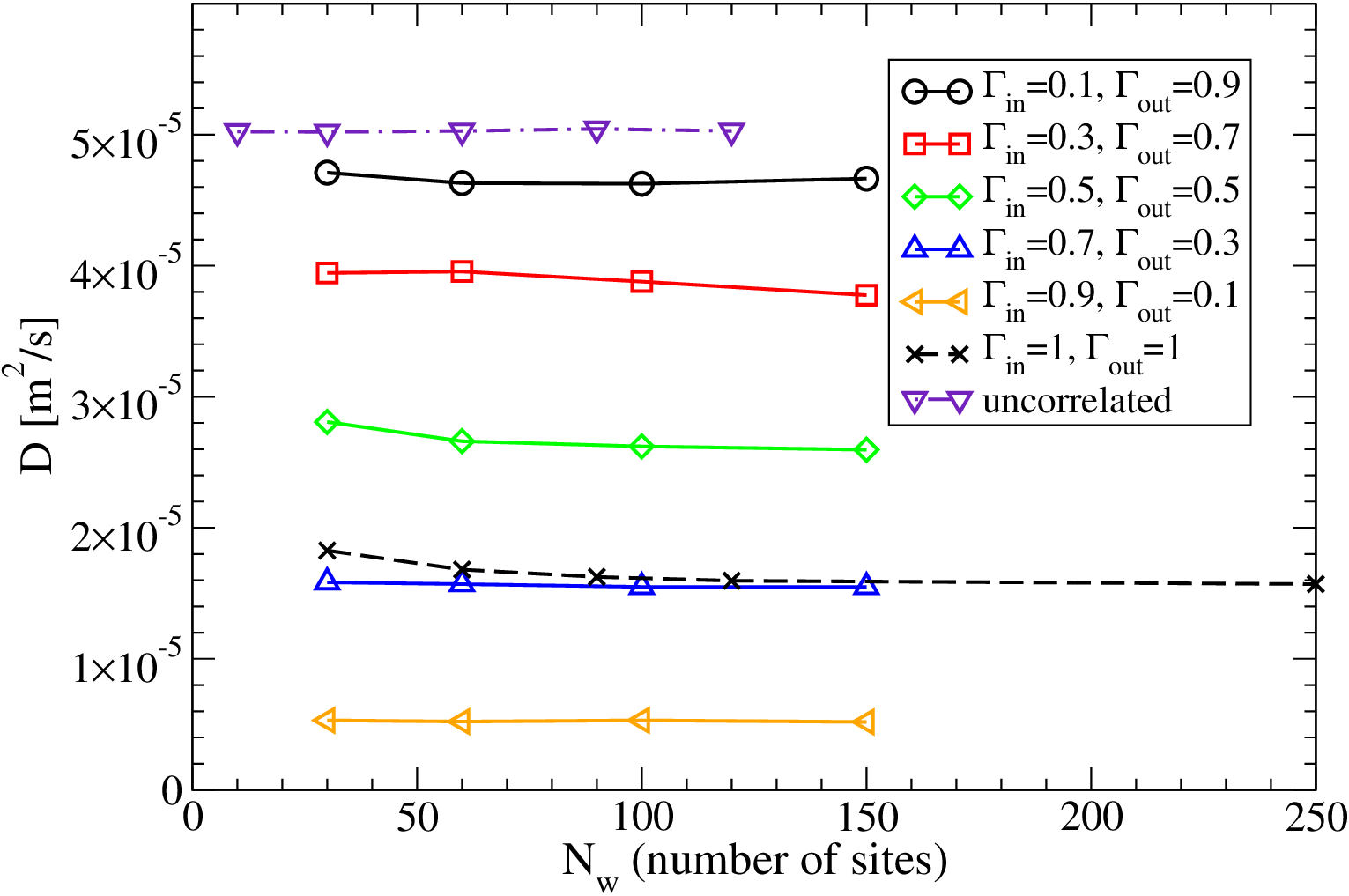}
\caption{ \label{f3}  (Color online) Linear chain. Spreading diffusion
  coefficient obtained from simulations  in  the presence of ASEP  and
  for  different input and output rates.   The  value for uncorrelated
  carriers is reported for comparison.  Curves are guide to the eyes.}
\end{figure}
\begin{figure}[t]
\centering
\includegraphics*[height=4.5cm]{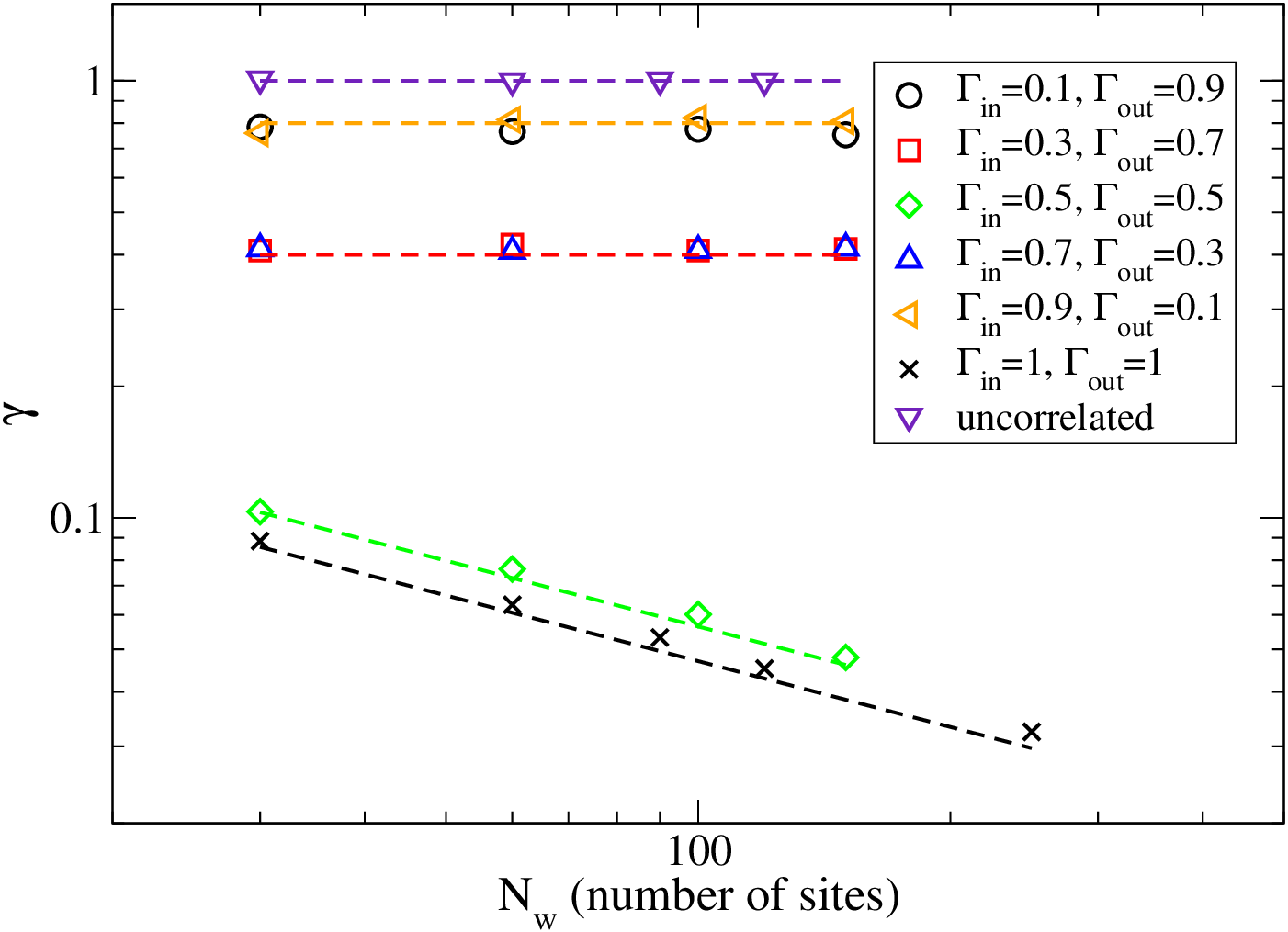}
\caption{  \label{f4} (Color online)  Linear chain. Comparison between
  analytical  Fano factors in  Eq. (\ref{gASEPl})  and full shot noise
  (dashed lines), with that obtained from simulations (symbols).}
\end{figure}

{\it B. Open linear chain.}  The  boundary conditions adequate to this
structure consists in connecting the two  terminals of the device with
two reservoirs,  where  $\Gamma_{in} \times \Gamma$  and $\Gamma_{out}
\times \Gamma$ are the transition rates from the left reservoir to the
first site of the device, and from the last site of  the device to the
right   reservoir, respectively.  For convenience,  the values of
$\Gamma_{in(out)}$
are  taken in the range between  0 and 1 being equivalent respectively
to   the      $\alpha$     and  $\beta$       parameters    in  
[\onlinecite{der1,der2}]     and   to     $f_L$   and    $(1-f_R)$    in
[\onlinecite{kor1}].
\par
Let us first consider the  case of  correlated carriers (ASEP  model).
For a given  carrier concentration   and  for large  $N_w$  analytical
theory  gives  for  the average  current  the same   expression of Eq.
(\ref{iASEP}).   The   diffusion  coefficient  of   the linear   chain
$\Delta_2$ (following [\onlinecite{der2}],
with the subscript $2$ labeling this specific case) takes the form: if
$\Gamma_{in} + \Gamma_{out}=1$,
\benl
\Delta_2=\left\{
\begin{array}{ll}
\frac{l^2\Gamma}{2}\Gamma_{in}\Gamma_{out} |\Gamma_{in}-\Gamma_{out}|
   & \textrm{when}\ \Gamma_{in} \neq \Gamma_{out} \\
\frac{l^2 \Gamma}{2} \frac{1}{4 (\pi N_w)^{1/2}}
   & \textrm{when}\ \Gamma_{in}=\Gamma_{out}
\end{array}
\right.
\ee
and, if $\Gamma_{in}=\Gamma_{out}=1$, 
\benl
\Delta_2 =\frac{l^2 \Gamma}{2} \frac{3 (2 \pi^{1/2})}{64 N_w^{1/2}}\ .
\ee
[Again $\Delta_2$ is supposed to differ in general
from $D$   obtained   from  Eq.(\ref{cdiff}).]   The  different  analytical
expressions for  $\Delta_2$ are a consequence
of the different  values taken by  $\Gamma_{in,out}$ and,  in turn, by
$\rho$ in the steady state.  Indeed,  the tuning  of the
$\Gamma_{in,out}$  controls the strength of the correlation among
carriers induced  by ASEP and thus  the particle density $\rho$ inside
the  device, as
summarized in the phase diagram reported in Fig. 2 of [\onlinecite{der1}].
\par
Figure \ref{f3} reports the  Fick's diffusion coefficient $D$ obtained
from simulations for the case of the linear chain.  Here, $D$ is found
to be practically independent of $N_w$.   Furthermore, in the presence
of ASEP   the  value of   $D$ is  systematically  lower  than  that of
uncorrelated particles $D_0$.  For the case $\Gamma_{in} +\Gamma_{out}
= 1$, the value of the diffusion  coefficient is well described by the
relation
\benl
D= \frac{l^2 \Gamma}{2} \Gamma_{out} = \frac{l^2 \Gamma}{2} (1-\rho)
\ee
We  notice that in the  above expression, $D$ becomes vanishing
small for $\Gamma_{out} \rightarrow 0$ because in this limit spreading
and current of carriers through the structure tends to stop.
In the  presence of ASEP the values of diffusion
obtained by simulations are found to differ significantly from those given
by analytical expressions, thus  implying that the quantities $\Delta_2$
an $D$ describe different microscopic processes.
\par
Figure \ref{f4} reports the results of the simulations for the current
noise [in this  case due to  the sum of all  the  contributions in Eq.
(\ref{SIdiv})] in terms of  the Fano factor.  From simulations, within
numerical uncertainty we find:
\be{conj}
S_I \equiv S_I^{v_d} + S_I^{N} + S_I^{Nv_d} + S_I^{v_dN} =
\frac{4e^2}{l^2} \Delta_2 \ .
\ee
with  the corresponding Fano factor satisfying the relation
\be{gASEPl}
\gamma=\frac{2 \Delta_2} {\Gamma l^2 \rho (1-\rho)}\ .
\ee
From the above expressions  we conclude that $\Delta_2$  describes the
total current noise instead of  the Fick's diffusion process.
The quantity $\Delta_2$
is found to  agree well with available analytical expressions,
as predicted by Eq.  (\ref{conj}),  in full agreement with the results
of [\onlinecite{der2}].  The  quantity  $D$
is found to depend  upon
the  degree of correlation, to be  independent of the  number of sites
and,    when  $\Gamma_{in}+\Gamma_{out}=1$,   to  be  proportional  to
$\Gamma_{out}$.   Since $\Delta_2$  is    related  to the  number   of
particles that  entered the   device  up to time  $t$, \cite{der2}  we
conclude that this  stochastic  quantity accounts  for both
velocity and number fluctuations.

\par
By  turning   off    the ASEP   correlation,    for  a given   carrier
concentration,  analytical theories give:  for the average current Eq.
(\ref{iFREE}), for the  diffusion coefficient Eq.  (\ref{dFREEr}), and
for the current noise (full shot noise is expected):
\benl
S_I \equiv S_I^{N}
= \frac{2e^2}{L^2} \langle v_d\rangle^2 \langle N\rangle \tau_{TR}
= 2e\langle I\rangle
\ee
which follows  from  a  correlation  function with   triangular  shape
vanishing at the transit time $\tau_{TR} = L/\langle v_d \rangle$.  As
a  consequence $\gamma=1$,    as confirmed by   the  results   of  the
simulations reported in Fig. \ref{f4}.

\begin{figure}[t]
\centering
\includegraphics*[height=4.5cm]{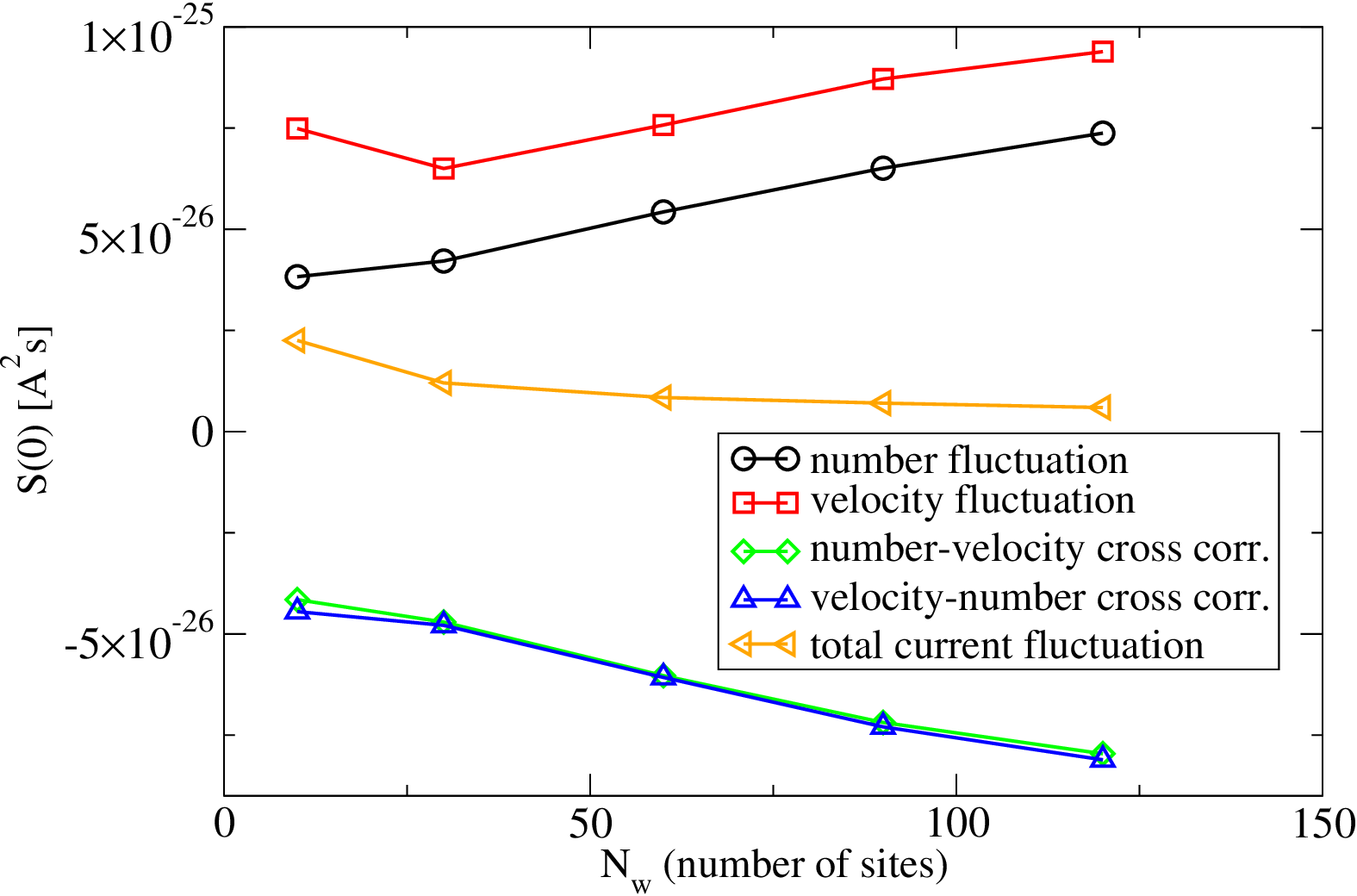}
\caption{ \label{f5} (Color online)  Different contributions and total
  noise power  for  the case   of   ASEP  in  the  open  linear   with
  $\Gamma_{in}=\Gamma_{out}=1$. Curves are guide to the eyes}
\end{figure}

\par
Finally, we  have investigated  separately   the contributions to  the
total  current  noise,  which  come  from velocity,  number, and their
cross-correlations in the presence of ASEP.  From the results
(\ref{f5}) we  can    see that  different contributions  are
comparable in magnitude, with  the cross terms, which  are responsible
of shot noise suppression, being negative.  We further notice that the
comparison between the two uncorrelated current-noise levels belonging
to the closed ring (Fig.  \ref{f2}) and to the open linear chain (Fig.
\ref{f3}) shows that  shot noise of  the linear chain exceeds velocity
noise of the closed ring for the ratio $L/l$, as expected.
\par
In conclusion,  we have carried out a  simulative investigation of the
inter-relations  between  noise   and   diffusion in  barrier  limited
transport under the ASEP condition.  For the case  of the closed ring,
since the number  of particles  is   fixed,
only the noise related to velocity  fluctuations is present.  
Here,   the diffusion
coefficient  obtained  from the  Fick's  law  is explicitly related
to current
noise, both  in    the presence and   in  the  absence of  the   ASEP.
Therefore,  evidence is  provided for  the  existence of a generalized
Nyquist-Einstein relation allowing the determination of diffusion from
a noise measurement or viceversa.  The correlations introduced by ASEP
are   found to be responsible  of  the dependence of  diffusion upon
 the
inverse  square root of the device  length.  For the  case of the open
linear chain the diffusion coefficient obtained from Fick's law is not
related to the current  noise, which now contains contributions coming
from  velocity,  number   and  their  cross-correlations.   Here   the
diffusion coefficient  is found to be independent  of  the number of
sites  but  to depend   on the strength   of  the correlation that  is
ultimately controlled by  the carrier density.    We remark, that  the
total current noise is found to be related to a ``diffusion constant''
(more properly a counting    statistics property \cite{der2}),  
whose analytical  expressions are satisfactorily confirmed    by
simulations.

Authors acknowledge valuable discussions  with Drs. V. Ya. Aleshkin, 
B. Derrida, and A. Korotkov.

\end{document}